\documentclass[onecolumn,preprint,aps,superscriptaddress,nofootinbib]{revtex4}
\usepackage{graphicx,amsmath}
\usepackage{hyperref}
\usepackage{amssymb}
\usepackage{color}
\usepackage[normalem]{ulem}

\newcommand{\beq}{\begin{equation}}
\newcommand{\eeq}{\end{equation}}
\newcommand{\bea}{\begin{eqnarray}}
\newcommand{\eea}{\end{eqnarray}}

\newcommand{\mev}{\mathrm{MeV}}
\newcommand{\cm}{\mathrm{cm}}

\newcommand{\ev}{\mathrm{eV}}
\newcommand{\gev}{\mathrm{GeV}}

\newcommand{\tev}{\mathrm{TeV}}

\newcommand{\nul}{\ensuremath{\nu}}
\newcommand{\pino}{\ensuremath{n}}
\newcommand{\nuh}{n}

\newcommand{\darkf}{\ensuremath{\psi}}
\newcommand{\cdarkf}{\ensuremath{\psi^c}}
\newcommand{\pili}{\ensuremath{\omega}}

\newcommand{\textprime}{\ensuremath{^\prime\,}}

\def\ct{c_\theta}
\def\st{s_\theta}
\def\tg{\tilde g}

\newcommand{\ignore}[1]{}

\makeatletter
\def\gsim{\mathrel{\lower2.5pt\vbox{\lineskip=0pt\baselineskip=0pt
           \hbox{$>$}\hbox{$\sim$}}}}
\def\lsim{\mathrel{\lower2.5pt\vbox{\lineskip=0pt\baselineskip=0pt
           \hbox{$<$}\hbox{$\sim$}}}}
\makeatother

\begin{document}
\setlength{\unitlength}{1mm}

\title{A Portalino to the Dark Sector}

\author{Martin Schmaltz}
\email{schmaltz@bu.edu}
\affiliation{Physics Department, Boston University, Boston, MA 02215}
\author{Neal Weiner}
 \email{neal.weiner@nyu.edu}
 \affiliation{Center for Cosmology and Particle Physics, Department of Physics, New York University, New York, NY 10003}

\begin{abstract}

``Portal'' models that connect the Standard Model to a Dark Sector allow for a wide variety of scenarios beyond the simplest WIMP models. Kinetic mixing of gauge fields in particular has allowed a broad range of new ideas. However, the models that evade CMB constraints are often non-generic, with new mass scales and operators to split states and suppress indirect detection signals. Models with a ``portalino'', a neutral fermion that marries a linear combination of a standard model neutrino and dark sector fermion and carries a conserved quantum number, can be simpler. This is especially interesting for interacting dark sectors; then the unmarried linear combination which we identify as the standard model neutrino inherits these interactions too, and provides a new, effective interaction between the dark sector and the standard model. These interactions can be simple $Z'$ type interactions or lepton-flavor changing. Dark matter freezes out into neutrinos, thereby evading CMB constraints, and conventional direct detection signals are largely absent. The model offers different signals, however. The ``portalino'' mechanism itself predicts small corrections to the standard model neutrino couplings as well as the possibility of discovering the portalino particle in collider experiments. Possible cosmological and astroparticle signatures include monochromatic neutrino signals from annihilation, spectral features in high energy CR neutrinos as well as conventional signals of additional light species and dark matter interactions.

\end{abstract}

\maketitle

\section{Searching for Hidden Sectors}
A major question for particle physics  is whether there is detectable physics beyond the standard model (BSM). We are well aware that there {\em is} physics beyond the standard model, as evidenced by dark matter, neutrino mass, gravity, and inflation. There are naturalness arguments in favor of additional BSM physics, such as the hierarchy problem and the strong CP problem. Since these connect to properties of known fields in the standard model, they often motivate interesting signals or new experiments. 

As the LHC energy has marched up and the luminosity increased, we have gained the ability to look for new particles at ever higher masses. Constraints on new particles with $O(0.1)$ level couplings are strong, with tremendous limits over wide ranges of lifetimes and properties. Simultaneously, attention has increasingly turned toward searches for new hidden sectors.

Much of the attention has come on ``dark sector'' models, where there can be new particles and interactions present, but which are generally assumed to be SM singlets. The communication between sectors occurs via ``portals,'' which are operators that connect the two sectors, i.e.,
\begin{equation}
	{\cal L} \supset \frac{O_{SM} O_{DS}}{\Lambda^p}.
\end{equation}
Where the dimension of the operator is $4+p$ and $\Lambda$ is the relevant scale of the operator. While non-renormalizable portals can be important, (see, e.g., \cite{Nomura:2008ru}), much effort has been focused on the renormalizable and super-renormalizable portals, namely the Higgs portals, the kinetic mixing portal, and the neutrino portal. While the Higgs portal typically yields WIMP-like models \cite{Burgess:2000yq} (although see \cite{Finkbeiner:2008qu} and related), the other two can yield scenarios with dramatically different mass ranges and properties.

The kinetic mixing portal, in particular, has received tremendous attention. In the context of a dark sector with charged matter and a dark Higgs, such kinetic mixing with a dark photon can naturally yield thermal dark matter over a wide range of scales \cite{Boehm:2003hm,Finkbeiner:2007kk,Pospelov:2007mp,ArkaniHamed:2008qn,Morrissey:2009ur, Essig:2013lka}. These models are simple and often yield interesting signals. Unfortunately, because of the coupling to charged particles the most straightforward of them also produce unobserved distortions of the CMB, absent new mass scales and operators to split the Dirac fermions, or to replace them with new scalar fields.

The neutrino portal has been well studied mostly in the context of sterile neutrino dark matter e.g., \cite{Dodelson:1993je,Asaka:2005an,Abazajian:2012ys}. However, it, too, can yield thermal models such as via a new particle for dark matter to annihilate into \cite{Bertoni:2014mva,Macias:2015cna,Gonzalez-Macias:2016vxy,Batell:2017rol,Campos:2017odj,Batell:2017cmf}. The right handed neutrino, given a Majorana mass, can also be integrated out, yielding non-renormalizable mass-mixing operators with charged dark sectors, e.g., \cite{Pospelov:2011ha}.  However, in these cases the light neutrino mass is corrected by an amount $\delta m_\nu \approx \sin^2 \theta m_{heavy}$, meaning either the mixing must be very small, or the heavy state must be $\ev$ in scale.

In this note, we show that the neutrino portal can also yield scenarios that are structurally as simple as kinetic mixing models, but instead have a dark sector that dominantly interacts with neutrinos, rather than charge. The effect arises from the inclusion of a ``portalino,'' a gauge-neutral fermion  with an exact or nearly exact global quantum number. The scenarios we arise at naturally have potentially sizable ($g_\nu \sim 10^{-2}$) new interactions for neutrinos, and dark matter freezeout into neutrinos. The dark matter in these scenarios can be light ($m_\chi \gsim 10\, \mev$), without needing to turn off annihilation channels because the only coupling to SM particles is to neutrinos which do not strongly affect the CMB. 

\section{The Portalino}
The neutrino portal is typically thought of as a relatively benign interaction. It produces a Dirac mass with a SM singlet fermion. If the singlet fermion has a large Majorana mass, the physical mass is suppressed by the seesaw mechanism and the Dirac mass can be sizable. If there is a lepton number symmetry, the  Dirac mass sets the scale for the neutrino mass and thus must be small enough to be consistent with terrestrial and cosmological measurements.

However, this latter case assumes that there are no other particles involved. It is this possibility that is our focus.

If one extends the model simply by adding a second singlet fermion \darkf\ the physical consequences are significant. If \darkf\ has a Dirac mass $m_n$ with the first singlet fermion, then there is a massless field in the spectrum. Specifically, there is a field we would identify as a neutrino 
\begin{equation}
	\nu  = \ct \nu_L + \st \darkf,
\end{equation}
with a massive partner
\begin{equation}
	\pino = \st \nu_L - \ct \darkf.
\end{equation}
Here $\tan \theta = m_D/m_n$, and the heavy mass eigenstate has mass $m = \sqrt{m_d^2 + m_n^2}$.
$\nu$ has its interactions suppressed compared to the standard model by $\ct$ which leads to strong constraints on the mixing angle $\st \lsim 10^{-1} -10^{-3}$ from precision measurements of weak decays with neutrinos. However, if the neutrino in question is $\nu_\tau$, there is far less precision information on its couplings and mixing angles as large as $\st\sim 0.3$ are possible. Importantly, the Dirac mass $m_d$ can naturally be as large as charged lepton masses, even if the heavy neutrino state is still weak scale.

Let us complicate the situation further - if we imagine the state \darkf\ has interactions of its own, whether scalar or vector, the light mass eigenstate will inherit those interactions, with a coupling suppressed by powers of $\st$. As will be our principal focus, let us suppose a coupling of \darkf\ to a new massive vector boson $\pili$
\begin{equation}
	{\cal{L}} \supset \darkf^\dagger \bar \sigma_{\mu} \darkf\, \pili^{\mu}.
\end{equation}
In terms of mass eigenstates, this interaction becomes
\begin{equation}
	{\cal{L}} \supset \st^2 \nul^\dagger \bar \sigma_{\mu} \nul \,\pili^{\mu}+\st \ct  \nul^\dagger \bar \sigma_{\mu} \nuh\, \pili^{\mu}+\st \ct \nuh^\dagger \bar \sigma_{\mu} \nul\, \pili^{\mu}+\ct^2 \nuh^\dagger \bar \sigma_{\mu} \nuh\, \pili^{\mu}.
\end{equation}
Thus, the light neutrino mass eigenstate, which we identify as the physical neutrino, carries a residual vector interaction as well. This ``effective Z\textprime '' process \cite{Fox:2011qd} is a simple way to add interactions to SM fermions. For other SM fermions, such effective Z\textprime$\!$ UV completions require new, light states charged with SM quantum numbers. The neutrino, uniquely, does not, and thus becomes a singular portal into new interactions of hidden sectors. 

Of course if \darkf\ is charged under a new gauge group, it cannot mix with a gauge singlet. If the gauge symmetry is broken, however, then, just as the standard model neutrino does, \darkf\ can marry the singlet, and the massless eigenstate will inherit its interactions. This singlet fermion which conveys these new interactions to the physical neutrino, and the resulting mass eigenstate we refer to as the ``portalino.''

The only theoretical requirement on this is that a suitably low-dimension operator exists which is a gauge singlet under the other gauge group and which creates a fermionic (fundamental or composite) single particle state. Unlike kinetic mixing which requires a $U(1)$ for a renormalizable interaction, here, regardless of how complicated the new sector's particle content is and what the gauge sector looks like, so long is there is a gauge singlet operator (akin to $L h$ in the SM) one can develop the interaction in question.

While this particular scenario is new, it draws from a number of ingredients that have existed in the literature. \cite{Fox:2011qd} discussed the effective Z\textprime\, scenario, the idea of using ``missing partners'' as a means to generate effective interactions with massive gauge bosons. \cite{Chang:2007de} discussed how with a missing partner, neutrinos with large Dirac masses can still have massless eigenstates, allowing the massless eigenstate to inherit a large Yukawa coupling. Interactions that only neutrinos feel, sometimes called ``secret'' interactions, have been widely discussed in many contexts \cite{Kopp:2014fha,Chu:2015ipa,Ng:2014pca}, often described as an effective theory.  Direct mass mixings of the form $\darkf \phi h l/M$ have been studied as a means to induce interactions for neutrinos \cite{Pospelov:2011ha, Pospelov:2012gm,Cherry:2014xra}, including in chiral models \cite{Berryman:2017twh}, but, absent tuning, these tend to require either small mixings or light sterile neutrinos. More closely related to dark matter, \cite{Farzan:2016wym} studied an effective Z\textprime model, where charged fermions marry the neutrino by extending the SM with a second Higgs doublet charged under a new $U(1)$. \cite{Bertoni:2014mva,Batell:2017cmf} showed how an inverse seesaw could yield a large $DM-\nu-\phi$ Yukawa interaction, where $\phi$ is some new lepton number carrying force carrier. These last two are closest in content to the scenario described here.

\subsection{A Simple Model}
\label{sec:simplest}

Taking the above discussion and translating into a full Lagrangian is straightforward - one must simply cancel gauge anomalies (by adding a conjugate $\cdarkf$) and ensuring that there are no additional massless states (which are constrained by the CMB).

The simplest example is
\bea
{\cal L} \supset y\, l h n^c +  y_n\, n^+\! \phi\, n^c + y_x\, x^-\! \phi^* x^c,
\label{eq:toylagrangian}
\eea  
where $l$ and $h$ are the usual Standard Model fields, $n^c$ and $x^c$ are gauge singlets and $n^{+}$ and $x^-$ are
two SM singlets which are charged under the dark $U(1)_d$. We use the $x$ and $n$ labels to distinguish the mass eigenstates after $U(1)_d$ breaking. Connecting to the previous section, $n^c$ is our portalino, $n^+ \leftrightarrow \darkf$, and we extend the model with additional fields to cancel the gauge anomalies and provide masses for all new states.
 Mass terms $n^c x^c$ and $n^+ x^-$ would be allowed by the gauge symmetries, but are easily forbidden with global symmetries. 

Assuming that the uneaten fields in the scalars $h$ and $\phi$ are heavy enough to be ignored we replace
them by their VEVs and we arrive at two (Dirac) mass terms
\bea
(y v_H \nu_L + y_n v_\phi\, n^+)n^c + y_x v_\phi\, x^- x^c = m_N (\sin \theta \nu_L + \cos \theta n^+)n^c + m_X x^- x^c \ .
\eea
There are two massive Dirac particles: $x^c$ pairs up with $x^-$, and $n^c$ combines with linear
combination ($n$) of $\nu_L$ and $n^+$, the other linear combination remains massless and is what is identified as 
the ``usual'' neutrino $\nu$. Note that we have included only one portalino field in this simple model. It couples to a linear combination of the SM lepton doublets $y l \rightarrow \sum_i y_i l_i$ in Eq.~\ref{eq:toylagrangian}. Generalizations with multiple portalinos are straightforward, see Section \ref{sec:generalization}.

The fields with couplings to gauge bosons in the mass eigenstate basis are
\bea
n\equiv \st\, \nu_L + \ct n^+, \qquad
\nu \equiv \ct\, \nu_L - \st n^+, \qquad
x\equiv x^- , 
\eea
with couplings
\bea
 W^+_\mu \quad &:& \quad \frac{g}{\sqrt{2}}\left[\ct (\nu^\dagger \bar \sigma^\mu  e)  + 
					\st (n^\dagger \bar \sigma^\mu  e) \right] \quad (+ \ h.c.\ {\rm for}\ W^-_\mu)\label{wcoup} \\
Z_\mu \quad &:& \quad \frac{\sqrt{g^2+g'^2}}{2}\left[\ct^2(\nu^\dagger \bar \sigma^\mu  \nu) + 
					 \st \ct(\nu^\dagger \bar \sigma^\mu  n + n^\dagger \bar \sigma^\mu  \nu) +
					 \st^2(n^\dagger \bar \sigma^\mu  n) \label{zcoup}\right] \\
\omega_\mu \quad &:&\quad \tg\left[ \st^2\,(\nu^\dagger \bar \sigma^\mu  \nu) + 
					 \st \ct\,(\nu^\dagger \bar \sigma^\mu  n + n^\dagger \bar \sigma^\mu  \nu) +
					 \ct^2\,(n^\dagger \bar \sigma^\mu  n) -
					 (x^\dagger \bar \sigma^\mu x) \right]\label{dphotoncoup} \ .
\eea

In this model, $n$ acts as an unstable heavy neutrino, the portalino, while $x$ is stable. As we shall discuss in Section \ref{sec:dm}, $x$ provides a natural dark matter candidate. The details of the model's dark phenomenology depend on the ordering of the massive particles, \pino, $x$ and $\pili$. All three obtain their masses from the symmetry breaking vev $v_\phi$ proportional to coupling constants. As a benchmark scenario we envision the $\pili$ as the heaviest particle with a mass of order GeV, the dark matter $x$ with a mass in the 10 to 100 MeV range and the portalino \pino\ somewhere in--between. With this ordering of the spectrum portalinos decay invisibly to dark matter particles, and dark matter freeze-out occurs via annihilation into neutrinos. Remarkably, all other orderings also produce viable models of thermal dark matter. 

\subsection{Non Abelian Models and generalizations}
\label{sec:generalization}

The $U(1)$ model is a very simple example of the portalino mechanism with an automatic dark matter candidate. However, there are a number of variants which we might also expect.

\subsubsection{Multiple Portalinos}
The simplest extension of the above scenario is to enlarge $n_P$, the total number of gauge singlet portalinos. As we do so, we should simultaneously consider enlarging $n_D$, the number of dark gauge-charged fermions $\darkf$. In the most straightforward examples $n_P = n_D$ and there are no new massless degrees of freedom. 

However, even in this simple modification, there are important differences. The first critical difference is that the heavier portalinos will dominantly decay invisibly, via an offshell $\pili$, irrespective of the ordering of the mass spectrum of \pino, $x$ and $\pili$. This will be critical as we discuss the experimental constraints on this scenario in Section~\ref{sec:bounds}.

The second difference is that the mixing angle of the lightest portalino is not necessarily the largest mixing angle. Thus, the massless ``neutrinos'' can have larger couplings to $\pili$, only constrained by the properties of a heavier, invisibly decaying portalino.

We can also consider moving away from $n_P = n_D$. If $n_P> n_D$, we will require some state to have extremely small ($< 10^{-12}$) Yukawas, lest the SM neutrinos have too large Dirac masses. Since our premise is that all these couplings should be more comparable to ordinary Yukawas, this moves the scenario in a qualitatively different direction.

If we take $n_D> n_P$, we will have new, massless degrees of freedom. As we will discuss later, there are well-known and generic constraints on new light degrees of freedom. There are unavoidable production processes for these states as well. Just as the portalino can decay to SM neutrinos, it will be able to decay to these new states as well. Four-fermi operators will allow production of these states via $\nu \nu \rightarrow \darkf \darkf$. If these states are charged under the $\omega$ gauge group, the rate of production will be large. If they are charged under some other gauge group, they could be produced by processes mediated by that gauge boson. However, even if those processes are suppressed, they will still be produced with a rate $T^5 \sin \theta_\nu^4 \sin \theta_\darkf^4/m_\pili^4$, where $\theta_\nu, \theta_\darkf$ are the mixings of $\nu$ and $\darkf$ with the portalino, respectively. Terminating this process by $T\sim \gev$, requires $m_\pili \gsim 10^5 \gev \times  \sin \theta_\darkf \sin \theta_\nu $. Precise constraints depend on the details of the new hidden sectors.

\subsubsection{Non-Abelian Models}
While the simplest model from Section~\ref{sec:simplest} is based on a $U(1)$ gauge group, the charged fields can be charged under any other group so long as a gauge singlet fermion operator is present. Simple extensions would be $SU(2)$ completely broken by a doublet,  $SU(2)\times U(1) \rightarrow U(1)$, akin to the standard model, or $SU(3) \rightarrow SU(2)$, broken by a triplet. Each of these illustrates interesting phenomenological differences.

For $G=SU(2)\rightarrow 0$, we would naturally require two portalinos to give masses to both components of the doublet \darkf. The off-diagonal $SU(2)$ interactions would mediate potential transitions between SM neutrinos. However, since the $\pili$ interactions in the single portalino case already needn't be flavor diagonal, we would not expect a significant change in neutrino properties. However dark matter is naturally a doublet as well, and there are a variety of interesting consequences if those states are non-degenerate. 

For $G=SU(2)\times U(1) \rightarrow U(1)$ there is naturally a charged partner, akin to the electron in the SM. The existence of the massless photon means that the theory should decouple before $T\sim \gev$ from the SM. One would expect multiple components of DM in this theory, at least one with a residual $U(1)$ interaction. 

For $G=SU(3)\rightarrow SU(2)$, $\darkf$ is a triplet. The portalino marries $\darkf_3$, while $\darkf_{1,2}$ remain charged under the unbroken $SU(2)$. With two copies of $\darkf$, one can write a non-vanishing $\epsilon^{ijk} \phi_i \psi_j \psi_k$, which would give masses to the $\darkf_{1,2}$ states. One could envision a component of the dark sector made of ``quirky'' dark matter. The variations of this scenario are large and we defer to later work.

\section{Dark Matter freezeout}
\label{sec:dm}

Since the field $\darkf$ is charged under a hidden gauge symmetry it is expected that there must be some additional field $\cdarkf$ to cancel the gauge anomalies. It is natural (although not required) that due to gauge or global charges of $\cdarkf$ no $\cdarkf \darkf$ Dirac mass term is present. If $\cdarkf$ acquires a mass by marrying a different field, it forms a natural candidate for dark matter.

This is simply illustrated within the context of the $U(1)$ model of Section~\ref{sec:simplest} where $x$ forms a natural dark matter candidate. 

The precise freezeout process depends on the spectrum. Depending on masses, $\chi \chi \rightarrow \nu \nu$, $\chi \chi \rightarrow \pino \pino$, $\chi \chi \rightarrow \pino \nu$ or $\chi \chi \rightarrow \pili \pili$ can all be the dominant annihilation channel. If $m_\chi > m_\pili$, then $\chi \chi \rightarrow \pili \pili$ typically dominates, with cross section $\sigma v = \pi \alpha^2/ m_\chi^2$.
 For light WIMPs, this requires a small coupling $\alpha \sim 10^{-4}\, [m_\chi/\gev]$. On the other hand, for $m_\chi < m_\pili$, s-channel annihilations are naturally mixing suppressed and couplings comparable to the SM are more naturally allowed. 

As a concrete example, let us choose $m_\pili= 1 \gev$, $m_\pino = 600 \mev$ and $m_X = 200 \mev$. In this case, the $\pili$ decays promptly to $\pino$ and $X$. The $\pino$ decays to $\nu X \bar X$ with width $g_D^4 \sin^2 \theta\, [m_\pino/m_\mu]^5\, [m_W/m_\pili]^4 \times \Gamma_\mu \sim g_D^4 \sin^2 \theta/(10^{-17} sec)$. 
Freezeout occurs when $\chi \chi \rightarrow \nu \nu$ decouples with $\sigma v \approx g_D^4 \sin^4 \theta m_\chi^2/m_\pili^4 = (g_D \sin \theta)^4/(5 \gev)^2$. We achieve the correct thermal cross section for dark matter freezeout with $(g_D \sin \theta)^2 \simeq 10^{-4}$.

\section{Portalino Phenomenology and Neutrino Constraints}
\label{sec:bounds}

As these scenarios can yield $O(1)$ corrections to the SM neutrino couplings, it is clear that constraints arise from a number of sources. Some constraints are modified in the presence of multiple portalinos, and we shall review them here. Our bounds are adapted from the excellent review by de Gouvea and Koblach \cite{deGouvea:2015euy}, the SHIP white paper \cite{Alekhin:2015byh}, and \cite{Bertoni:2014mva,Batell:2017cmf}.

\begin{figure}[!htbp]%
	\centering
	\includegraphics[width=0.49\textwidth]{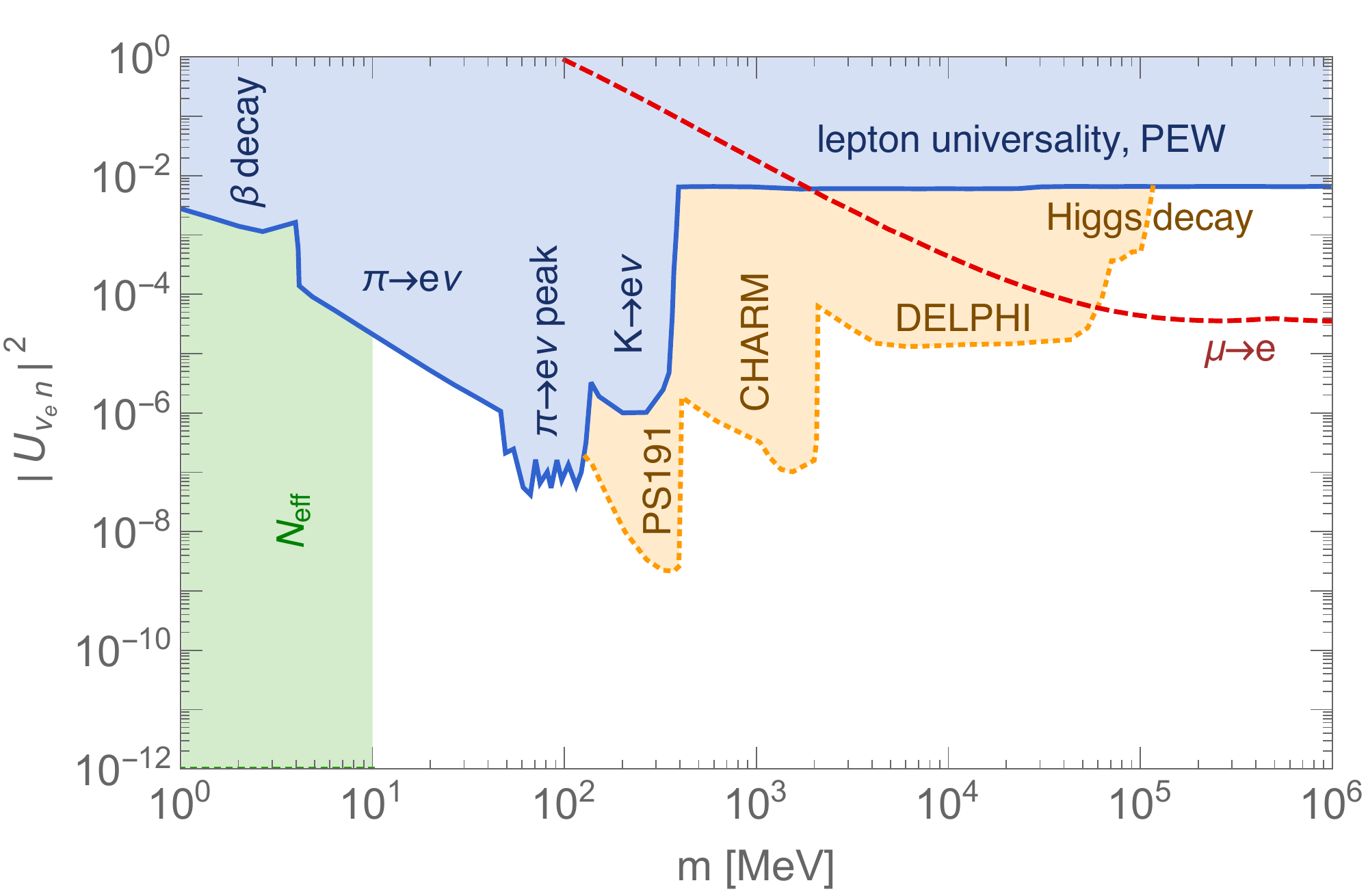}
	\includegraphics[width=0.49\textwidth]{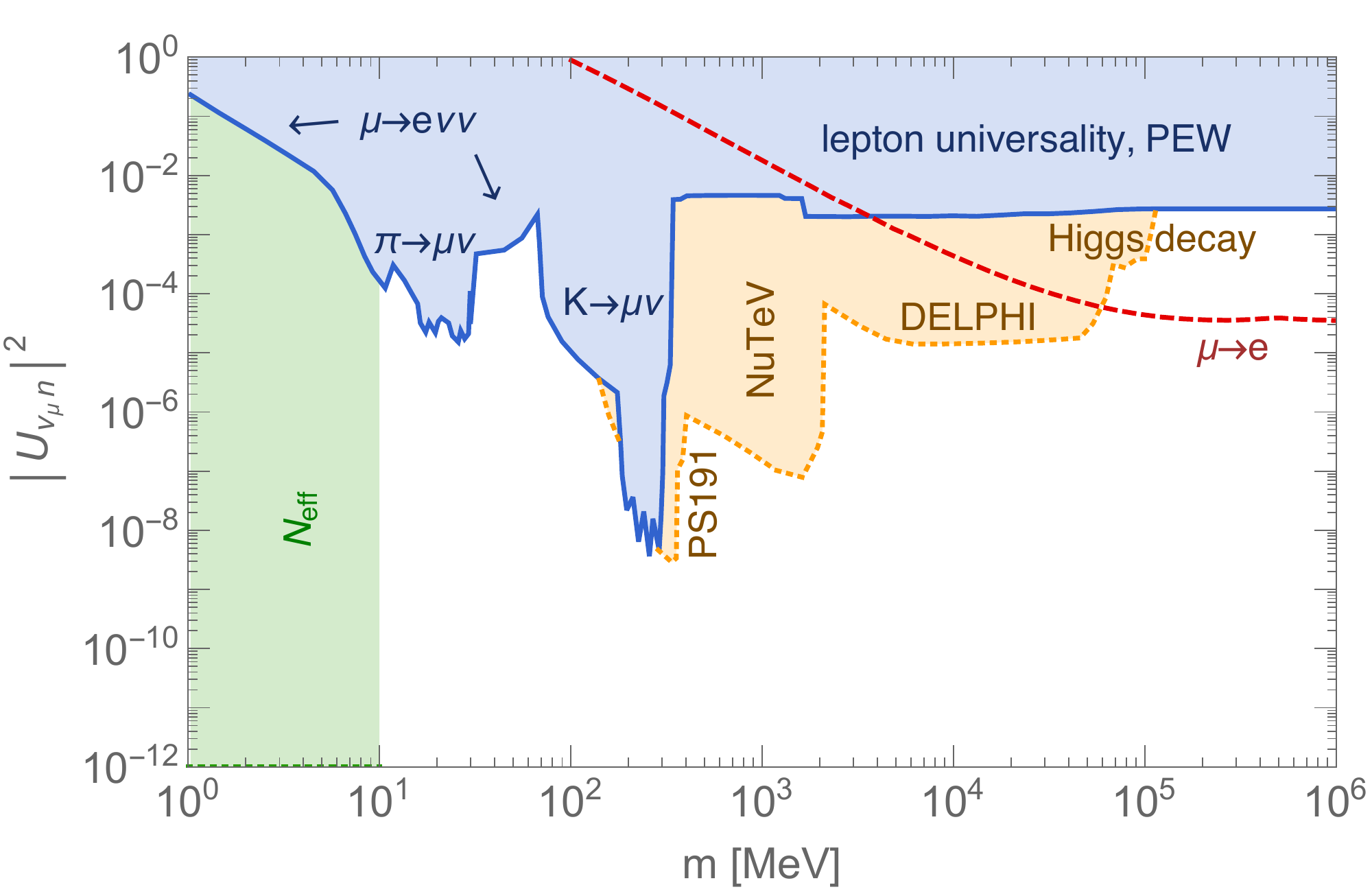}
         \includegraphics[width=0.49\textwidth]{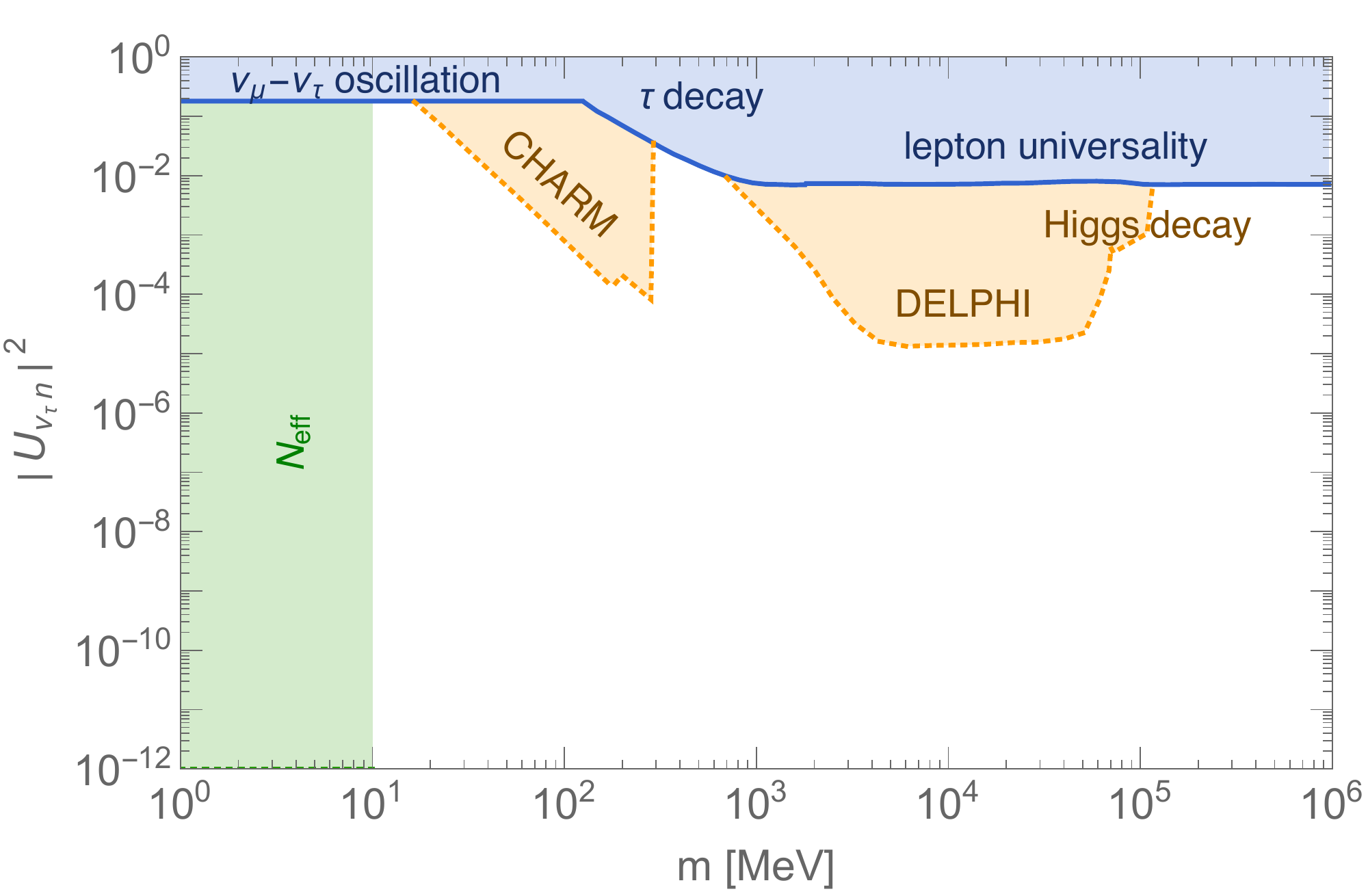}
\caption{Bounds in the mass versus mixing parameter space for portalinos mixing with electron neutrinos ({\it left}), muon neutrinos ({\it right}), and tau neutrinos ({\it bottom}).
In all three plots portalino masses smaller than 10 MeV are ruled out by the CMB bound on extra relativistic degrees of freedom $\Delta N_{eff}$ (green shading). The blue shaded regions at large mixing angles are ruled out from the combination of a number of fairly model-independent constraints which do not rely on portalino decays. These including bounds on non-universal neutrino couplings, precision electroweak constraints, changes in the rates and kinematics of $\pi$, K, $\mu$ and $\tau$ decays,
and neutrino oscillations. The dotted line bounds (orange shading) are more model dependent as they rely on portalino decays to visible SM particles. For details, see text and references \cite{deGouvea:2015euy, Deppisch:2015qwa, Alekhin:2015byh, Bertoni:2014mva}.}%
\label{fig:bounds}\end{figure}

\begin{enumerate}
\item 
{\it Precision electroweak and lepton universality}: In the Standard Model the muon width is proportional to the Fermi constant $G_F$ squared which can also be determined in other ways, for example by measuring the $W$ and $Z$ masses and $\alpha_{em}$. Comparing independent determinations of $G_F$ is a test of the SM. A portalino mixing with either the electron or muon neutrino reduces the $W$ coupling of the neutrino by $\cos \theta$. This reduces the muon width by $\cos^2 \theta$ and comparing to other experimental determinations of $G_F$ one obtains an upper bound on the angle $\theta$ as shown in Fig.~\ref{fig:bounds}, labeled ``lepton universality, PEW''. Muon decay bounds become weaker if the portalino mass is lower than the muon mass so that portalino final states are possible.  In the limit $m_\pino \ll m_\mu$, the muon width including neutrino and portalino final states is proportional to  $\cos^2 \theta + \sin^2 \theta =1$, i.e. insensitive to $\theta$. For non-negligible portalino masses a phase space analysis of Michel electrons (labeled $\mu \to e\nu\nu$ in Fig.~\ref{fig:bounds}) can give a strong bound. Similarly, $\tau$ decays can be used to bound mixing of the portalino with $\tau$ neutrinos. 

\item
{\it Meson decays}: Charged current meson decays with leptons in the final state can go to portalinos if the portalino is lighter than the decaying meson. In the case of two-body decays of stopped mesons such as $\pi \rightarrow e \nu$ or $K \rightarrow e \nu$ the energy of the charged lepton is monochromatic and would be shifted to lower values for decays with portalinos. The absence of a second line in the spectrum of final state charged lepton energies provides a very strong constraint because of the large number of pion and Kaon decays observed. In addition, the overall rates for leptonic $K$ and $\pi$ decays are sensitive to mixing with portalinos. 
In hadronic $\tau$ decays the phase space distribution of final state hadrons is sensitive to the presence of a final state portalino with non-negligible mass. 

\item
{\it Neutrino oscillations}: Another bound on portalino-neutrino mixing which does not rely on details of the portalino decay can be obtained by considering neutrino oscillations. If neutrinos mix significantly with a sterile portalino then the observed $3\times 3$ neutrino mixing matrix is non-unitary. An analysis of atmospheric neutrino data gives the bound $|U_{\nu_\tau n}|^2 \lsim 0.18$. 

\item
{\it Portalino decays to visible final states}: The previous types of portalino searches do not require observation of the portalino itself. They rely on the difference in phase space distributions of the other particles produced in decays. If the portalino itself is unstable on detector time scales and decays to visible decay products then one can look for the decays of portalinos. For example, portalinos may be produced in rare $B$ decays or in $Z$ decays with subsequent prompt portalino decays to charged leptons. Such decays have been searched for and provide stringent bounds for heavier portalinos which decay promptly to visible decay products. We indicate such bounds in Figures \ref{fig:bounds} with dotted lines bounding orange regions. In addition, if portalinos are sufficiently long-lived they will have displaced decays. Such portalinos could be produced in high luminosity beam dump experiments where they pass through shielding material before decaying in a detector cavity. Experiments have a significant reach even in the case of very small mixing but only when the portalino decays into visible final states. We also indicate such bounds from in Figure \ref{fig:bounds} with dotted lines. Significant improvement of such bounds would be obtained with SHIP \cite{Bonivento:2013jag,Alekhin:2015byh} LBNL \cite{Adams:2013qkq}, and FCC-ee \cite{Blondel:2014bra}.

\item
{\it Lepton flavor violation}: The portalino can mix with more than one flavor of neutrino and therefore mediate lepton flavor changing transitions. In the SM, the corresponding neutrino-mediated transitions are negligible because of the smallness of the neutrino masses, but here the Dirac mass terms of the portalino are much larger. Currently, the only bounds which are competitive with flavor-preserving bounds are from $\mu\rightarrow e$ transitions and bound the product
$|U_{\nu_\mu n} U_{\nu_e n}| \sim \sin \theta_{\mu}\, \sin \theta_e$. The bounds shown in Figure \ref{fig:bounds} (red dashed lines) assume that the portalino mixes equally strongly with $\nu_\mu$ and $\nu_e$, i.e. $\sin \theta_{\mu} = \sin \theta_e$.

\end{enumerate}

\section{Portalino Cosmology}

The portalino, by virtue of its relatively large interactions with the SM, is in thermal and chemical equilibrium in the early universe and does not fall out of equilibrium until temperatures below its mass where its number density becomes exponentially suppressed. To see this consider the rate for annihilation $\pino \pino \rightarrow \pino \nu$ at temperatures near the portalino mass $T\sim m_\pino$
\bea
\Gamma_{ann} &\sim& n_n \sigma v \sim m_\pino^3\ \frac{m_\pino^2 g_D^4 \sin^2\theta }{m_\pili^4} \\
&\sim& 10^{-12}\, \mev \ \left[\frac{m_\pino}{10\mev}\right]^5 \left[\frac{\gev}{m_\pili}\right]^4
\left[\frac{g_D^4 \sin^2\theta}{10^{-5}}\right] 
\label{eq:annrate}
\eea
which is much larger than the Hubble rate $H\sim m_\pino^2/M_{pl} \sim 10^{-19}\, \mev\ [m_\pino/10\mev]^2$ even for portalinos as light as 10 $\mev$. At lower temperatures, it falls out of equilibrium and we check that its lifetime is short compared with the time of BBN. The lifetime depends on whether it decays into dark sector or SM states.  If the portalino is the lightest dark sector particle, it can still decay as $\pino \rightarrow 3 \nu$ via an off-shell $\pili$ with
\bea
\Gamma_\pino &\approx& \frac{1}{10^{-2} \sec} \  \left[\frac{m_\pino}{10 \mev}\right]^5 \left[\frac{\gev}{m_\pili}\right]^4
\left[\frac{g_D^4\sin^2 \theta}{10^{-5}}\right]  \left[\frac{\sin^4 \theta_{max}}{10^{-6}}\right].
\eea
Thus even for portalinos which can only decay to SM particles the decay is sufficiently rapid. When there is a dark sector state that is lighter than the portalino under consideration (for example, if there is a lighter portalino or if the dark matter particle is lighter than the portalino) then the lifetime for three body decays is much shorter with $\sin^2 \theta_{max} \rightarrow 1$. If the portalino is heavier than $m_\pili$ then the two body decay $\pino \rightarrow \pili \nu$ becomes dominant.
Thus even portalinos as light as 10 $\mev$ can decay promptly before BBN under a wide range of circumstances.

In determining the 10 $\mev$ lower bound for the portalino mass, the dominant constraint comes from CMB bounds on light species. 
The portalino generally stays in kinetic and chemical equilibrium until after the neutrinos decouple from the electron/photon bath. The total portalino entropy at neutrino decoupling is then ultimately deposited into the neutrino bath and increases $N_{eff}$. The temperature of chemical decoupling for the different neutrinos is approximately $T^{chem}_{\nu_e,\nu_{\mu},\nu_{\tau}} \simeq 3.2\, \mev, 5.3\,\mev, 5.3\, \mev$ \cite{Dolgov:2002wy}.  We take a limit $ N_{eff} < 3.37$ arising from a combination of Planck and other data \cite{Ade:2015xua}. This results in a bound of $m_\pino > 22, 36\, \mev$ assuming the portalino can annihilate to $e$ or only $\mu/\tau$, respectively. Even with a small coupling to $\nu_e$ the lower bound typically applies (see Eq.~\ref{eq:annrate}).  Both of these bounds assume that the portalinos cannot deposit their energy directly into SM particles other than the neutrinos, and, in the latter case, that the $\nu_{\mu,\tau}$ cannot rethermalize with $\nu_e$ via the new $\pili$ interactions (in which case the lower $\nu_e$ bound applies). Since this analysis assumes an instantaneous decoupling of neutrinos from the electron/photon bath, we conservatively plot a bound of $m_\pino > 10\,\mev$ in Fig.~\ref{fig:bounds}.\footnote{Note that if there is a very late phase transition, a light sterile state is allowable \cite{Vecchi:2016lty}, but this yields qualitatively different phenomenology.}

\subsection{Light Species}
In this scenario, it is quite common that there are additional light species present. 
This can be because the gauge sector has a residual unbroken component, or non-Abelian partners of \darkf\ have small or zero masses. Thus, it is worth considering what constraints on $N_{eff}$ imply for such portalino scenarios.

Assuming that the hidden sector decouples from the SM at a temperature $T_{dec}$, then the effective number of neutrinos contributed by the particles in the hidden sector is

\begin{equation}
	\Delta N_{eff} = \frac{4}{7}\, g_{*D} \left(\frac{g_{*D}^{dec}\, g_{* SM}}{g_{* D}\, g_{* SM}^{dec} }\right)^{4/3},
\end{equation}
where $g_{* D}, g_{* SM}, g_{*D}^{dec}, g_{* SM}^{dec}$ are the effective number of degrees of freedom in the dark sector and the SM at low energies, and in the dark sector and the SM at decoupling, respectively. Specifically, $g_{* SM}= 10.75$, since we are anchoring to the last point when neutrinos have their entropy increased. For simplicity, we can take $g_{* D}= g_{*D}^{dec}$ and assume $T_{dec}\gsim 1 \gev$ at which time $g_{* SM}^{dec} = 61.75$, yielding
\begin{equation}
	\Delta N_{eff} = .056 g_{* D}.
\end{equation}
Taking the Planck limit (95\% confidence) of $\Delta N_{eff}  < 0.33$ \cite{Ade:2015xua} we find the relatively mild $g_{* D} < 6$. On the other hand, If $T_{dec}\lsim \Lambda_{QCD}$ at which time $g_{* SM}^{dec} = 17.25$, this becomes a more stringent requirement $g_{* D} < 1.1$.

\section{Discussion}
Dark matter that primarily interacts with neutrinos is a challenging scenario to test. Mixing of the portalino with SM neutrinos is a crucial component of the scenario and can be tested by precision measurements of the SM neutrino couplings. But there are also several more model-dependent possible signals which depend on the details of the hidden sector. 

\begin{itemize}    

\item
In addition to the large interaction of the $\pili$ with neutrinos, it may have small interactions with other SM particles. This could arise, for instance, from a small kinetic mixing with the SM photon. Alternatively, we could identify it as the gauge boson of some other SM symmetry, such as baryon number, $\mu-\tau$, or some effective $Z\textprime\!$, although this would require either small couplings or $m_\pili \gsim 100 \gev$. In addition to the well-studied phenomena associated with those forces, this would also yield signals of enhanced $\nu-SM$ interactions, especially at energies comparable to $m_\pili$.

\item
The portalino couplings may also violate lepton flavor. Then $W$ loops in association with flavor violating portalino coupling insertions can give rise to processes like $\mu \rightarrow e \gamma$ and $\mu$ to $e$ conversion in the background of a nucleus. The Mu2e experiment \cite{Bartoszek:2014mya} at Fermilab will look for $\mu$ to $e$ conversion in the field of an Aluminum nucleus, and is expected to improve the limits on $|U_{\nu_\mu n} U_{\nu_e n}|$ by two orders of magnitude.

\item
A clear signature of this scenario would be the detection of  monochromatic neutrinos from dark matter annihilation. Current limits on this from the galactic center are roughly $100 - 1000$ times thermal from $10\, \gev \lsim m_\chi \lsim \tev$ \cite{Adrian-Martinez:2015wey,Aartsen:2016pfc}, making a straightforward detection of the scenarios described above challenging. However, it is quite simple to employ the portalino to yield models that could be detected. 

In particular, one can envision a scenario with $m_\pili < 2 m_\pino$ and dark matter a vectorlike state with $\sim \tev$ mass, charged under $\pili$. This allows one to straightforwardly adopt the construction of \cite{ArkaniHamed:2008qn}, and consider $\chi \chi \rightarrow \omega \omega$, with $\omega \rightarrow \nu \bar \nu$. For $m_\pili \sim \gev$ a sizable Sommerfeld enhancement would boost the signal into the detectable regime. For $m_\pili> 2 m_\pino$ the signals could become partially visible at the level of the visible BR of the $\pino$, yielding other signatures, including CMB constraints. 

Finally, superheavy dark matter could conceivably decay via $\omega$ emission. The boosted $\omega$ could then decay producing ultra high energy neutrinos, but without any associated charged particle signals, evading the basic constraints considered in \cite{Cohen:2016uyg}.

\item
Another possible signal would be on the spectrum of UHE neutrinos observed at IceCube. The center of mass energy for a PeV cosmic ray neutrino incident on a non-relativistic neutrino of the relic neutrino background with mass $O(0.1\ev)$ is $O(100\, \mev)$. Thus, it is an intriguing point that the ongoing search at IceCube is for the first time giving us information on $\nu-\nu$ interactions in the $10\,\mev - 1\, \gev$ range. 

Given this, it is conceivable to consider a $\pili$-burst scenario akin to the Z-burst idea \cite{Weiler:1997sh}. The average density of relic neutrinos is $O(100\, \cm^{-3})$, thus, we can consider the column density for a neutrino traversing the observable universe,
\begin{equation}
	c \tau n_{\nu} \approx (3\times10^{10} cm/sec)(5\times10^{17} sec) (100/cm^3) \approx 10^{30} /cm^2 \approx (20 \gev)^2.
\end{equation}
Thus, for $\sigma \sim \sin^4 \theta/m_\pili^2 \sim (10^{-3})^2/(100 \mev)^2 \sim (100\,\gev)^{-2}$ (which is the approximate size of the cross section on resonance), one can reasonably have a universe that is somewhat opaque to neutrinos at the resonance energy. This would lead to distortions of the cosmic ray neutrino spectrum which could be detectable at IceCube.\cite{Ng:2014pca,Ioka:2014kca}

\item
Self-interactions of the dark matter may thermalize the cores of dark matter halos, potentially resolving small scale anomalies \cite{Spergel:1999mh,Loeb:2010gj,Tulin:2013teo,Kaplinghat:2015aga}. The interactions between the dark matter and neutrinos or between dark matter and dark radiation in our models can also leave a measurable imprint on the large scale structure of the universe \cite{Boehm:2000gq,Buen-Abad:2015ova}

\end{itemize}

The portalino could naturally find itself embedded in other scenarios, such as solutions to the hierarchy problem. In a SUSY model, the mass scale could arise radiatively, analogously to kinetic mixing scenarios \cite{Hooper:2008im,ArkaniHamed:2008qp,Cheung:2009qd}. In a Twin Higgs scenario,  portalinos could serve as a means to marry off just three of the six neutrinos, ameliorating the massless degree of freedom problem in those models. We leave a detailed study of these possibilities for future work.

We should also note that while we have focused on the portalino coupling to the neutrino and carrying an effective lepton number, it is also possible to write the non-renormalizable operator $u d d \pino/M^2$, replacing lepton number with neutron number. This would yield a small $\pino$-neutron mass mixing, and by analogy with the neutrino mixing scenario, would lead to neutron-specific effective interactions. Given the tremendous questions of flavor and collider limits, a full discussion warrants further study.

The astute reader will have noticed that we have not said anything about neutrino masses. This is because the portalino is compatible with many different ideas for neutrino mass generation. One scenario that appears particularly intriguing is radiative neutrino mass generation via \pino\ number violation. Or one could envision an inverse seesaw due to a small Majorana mass for \darkf.

While we have illustrated a variety of interesting scenarios, we have only scratched the surface of possible models. In particular, chiral models, non-Abelian models, and a thorough exploration of dark matter scenarios is warranted.

\section*{Acknowledgments}
\label{sec:ack}

We thank Andre de Gouvea and Little Dave for helpful comments and discussions. The work of MS is supported by DOE grant DE-SC0015845. The work of NW is supported by the NSF under grant PHY-1620727. We would like to thank the Aspen Center for Physics, which is supported by NSF grant PHY-1607761, for hospitality during work on this project.

\bibliography{portabib}

\begin{thebibliography}{50}
\expandafter\ifx\csname natexlab\endcsname\relax\def\natexlab#1{#1}\fi
\expandafter\ifx\csname bibnamefont\endcsname\relax
  \def\bibnamefont#1{#1}\fi
\expandafter\ifx\csname bibfnamefont\endcsname\relax
  \def\bibfnamefont#1{#1}\fi
\expandafter\ifx\csname citenamefont\endcsname\relax
  \def\citenamefont#1{#1}\fi
\expandafter\ifx\csname url\endcsname\relax
  \def\url#1{\texttt{#1}}\fi
\expandafter\ifx\csname urlprefix\endcsname\relax\def\urlprefix{URL }\fi
\providecommand{\bibinfo}[2]{#2}
\providecommand{\eprint}[2][]{\url{#2}}

\bibitem[{\citenamefont{Nomura and Thaler}(2009)}]{Nomura:2008ru}
\bibinfo{author}{\bibfnamefont{Y.}~\bibnamefont{Nomura}} \bibnamefont{and}
  \bibinfo{author}{\bibfnamefont{J.}~\bibnamefont{Thaler}},
  \bibinfo{journal}{Phys. Rev.} \textbf{\bibinfo{volume}{D79}},
  \bibinfo{pages}{075008} (\bibinfo{year}{2009}), \eprint{0810.5397}.

\bibitem[{\citenamefont{Burgess et~al.}(2001)\citenamefont{Burgess, Pospelov,
  and ter Veldhuis}}]{Burgess:2000yq}
\bibinfo{author}{\bibfnamefont{C.~P.} \bibnamefont{Burgess}},
  \bibinfo{author}{\bibfnamefont{M.}~\bibnamefont{Pospelov}}, \bibnamefont{and}
  \bibinfo{author}{\bibfnamefont{T.}~\bibnamefont{ter Veldhuis}},
  \bibinfo{journal}{Nucl. Phys.} \textbf{\bibinfo{volume}{B619}},
  \bibinfo{pages}{709} (\bibinfo{year}{2001}), \eprint{hep-ph/0011335}.

\bibitem[{\citenamefont{Finkbeiner et~al.}(2008)\citenamefont{Finkbeiner,
  Slatyer, and Weiner}}]{Finkbeiner:2008qu}
\bibinfo{author}{\bibfnamefont{D.~P.} \bibnamefont{Finkbeiner}},
  \bibinfo{author}{\bibfnamefont{T.~R.} \bibnamefont{Slatyer}},
  \bibnamefont{and} \bibinfo{author}{\bibfnamefont{N.}~\bibnamefont{Weiner}},
  \bibinfo{journal}{Phys. Rev.} \textbf{\bibinfo{volume}{D78}},
  \bibinfo{pages}{116006} (\bibinfo{year}{2008}), \eprint{0810.0722}.

\bibitem[{\citenamefont{Boehm and Fayet}(2004)}]{Boehm:2003hm}
\bibinfo{author}{\bibfnamefont{C.}~\bibnamefont{Boehm}} \bibnamefont{and}
  \bibinfo{author}{\bibfnamefont{P.}~\bibnamefont{Fayet}},
  \bibinfo{journal}{Nucl. Phys.} \textbf{\bibinfo{volume}{B683}},
  \bibinfo{pages}{219} (\bibinfo{year}{2004}), \eprint{hep-ph/0305261}.

\bibitem[{\citenamefont{Finkbeiner and Weiner}(2007)}]{Finkbeiner:2007kk}
\bibinfo{author}{\bibfnamefont{D.~P.} \bibnamefont{Finkbeiner}}
  \bibnamefont{and} \bibinfo{author}{\bibfnamefont{N.}~\bibnamefont{Weiner}},
  \bibinfo{journal}{Phys. Rev.} \textbf{\bibinfo{volume}{D76}},
  \bibinfo{pages}{083519} (\bibinfo{year}{2007}), \eprint{astro-ph/0702587}.

\bibitem[{\citenamefont{Pospelov et~al.}(2008)\citenamefont{Pospelov, Ritz, and
  Voloshin}}]{Pospelov:2007mp}
\bibinfo{author}{\bibfnamefont{M.}~\bibnamefont{Pospelov}},
  \bibinfo{author}{\bibfnamefont{A.}~\bibnamefont{Ritz}}, \bibnamefont{and}
  \bibinfo{author}{\bibfnamefont{M.~B.} \bibnamefont{Voloshin}},
  \bibinfo{journal}{Phys. Lett.} \textbf{\bibinfo{volume}{B662}},
  \bibinfo{pages}{53} (\bibinfo{year}{2008}), \eprint{0711.4866}.

\bibitem[{\citenamefont{Arkani-Hamed et~al.}(2009)\citenamefont{Arkani-Hamed,
  Finkbeiner, Slatyer, and Weiner}}]{ArkaniHamed:2008qn}
\bibinfo{author}{\bibfnamefont{N.}~\bibnamefont{Arkani-Hamed}},
  \bibinfo{author}{\bibfnamefont{D.~P.} \bibnamefont{Finkbeiner}},
  \bibinfo{author}{\bibfnamefont{T.~R.} \bibnamefont{Slatyer}},
  \bibnamefont{and} \bibinfo{author}{\bibfnamefont{N.}~\bibnamefont{Weiner}},
  \bibinfo{journal}{Phys. Rev.} \textbf{\bibinfo{volume}{D79}},
  \bibinfo{pages}{015014} (\bibinfo{year}{2009}), \eprint{0810.0713}.

\bibitem[{\citenamefont{Morrissey et~al.}(2009)\citenamefont{Morrissey, Poland,
  and Zurek}}]{Morrissey:2009ur}
\bibinfo{author}{\bibfnamefont{D.~E.} \bibnamefont{Morrissey}},
  \bibinfo{author}{\bibfnamefont{D.}~\bibnamefont{Poland}}, \bibnamefont{and}
  \bibinfo{author}{\bibfnamefont{K.~M.} \bibnamefont{Zurek}},
  \bibinfo{journal}{JHEP} \textbf{\bibinfo{volume}{07}}, \bibinfo{pages}{050}
  (\bibinfo{year}{2009}), \eprint{0904.2567}.

\bibitem[{\citenamefont{Essig et~al.}(2013)}]{Essig:2013lka}
\bibinfo{author}{\bibfnamefont{R.}~\bibnamefont{Essig}} \bibnamefont{et~al.},
  in \emph{\bibinfo{booktitle}{{Proceedings, 2013 Community Summer Study on the
  Future of U.S. Particle Physics: Snowmass on the Mississippi (CSS2013):
  Minneapolis, MN, USA, July 29-August 6, 2013}}} (\bibinfo{year}{2013}),
  \eprint{1311.0029},
  \urlprefix\url{http://inspirehep.net/record/1263039/files/arXiv:1311.0029.pdf}.

\bibitem[{\citenamefont{Dodelson and Widrow}(1994)}]{Dodelson:1993je}
\bibinfo{author}{\bibfnamefont{S.}~\bibnamefont{Dodelson}} \bibnamefont{and}
  \bibinfo{author}{\bibfnamefont{L.~M.} \bibnamefont{Widrow}},
  \bibinfo{journal}{Phys. Rev. Lett.} \textbf{\bibinfo{volume}{72}},
  \bibinfo{pages}{17} (\bibinfo{year}{1994}), \eprint{hep-ph/9303287}.

\bibitem[{\citenamefont{Asaka et~al.}(2005)\citenamefont{Asaka, Blanchet, and
  Shaposhnikov}}]{Asaka:2005an}
\bibinfo{author}{\bibfnamefont{T.}~\bibnamefont{Asaka}},
  \bibinfo{author}{\bibfnamefont{S.}~\bibnamefont{Blanchet}}, \bibnamefont{and}
  \bibinfo{author}{\bibfnamefont{M.}~\bibnamefont{Shaposhnikov}},
  \bibinfo{journal}{Phys. Lett.} \textbf{\bibinfo{volume}{B631}},
  \bibinfo{pages}{151} (\bibinfo{year}{2005}), \eprint{hep-ph/0503065}.

\bibitem[{\citenamefont{Abazajian et~al.}(2012)}]{Abazajian:2012ys}
\bibinfo{author}{\bibfnamefont{K.~N.} \bibnamefont{Abazajian}}
  \bibnamefont{et~al.} (\bibinfo{year}{2012}), \eprint{1204.5379}.

\bibitem[{\citenamefont{Bertoni et~al.}(2015)\citenamefont{Bertoni, Ipek,
  McKeen, and Nelson}}]{Bertoni:2014mva}
\bibinfo{author}{\bibfnamefont{B.}~\bibnamefont{Bertoni}},
  \bibinfo{author}{\bibfnamefont{S.}~\bibnamefont{Ipek}},
  \bibinfo{author}{\bibfnamefont{D.}~\bibnamefont{McKeen}}, \bibnamefont{and}
  \bibinfo{author}{\bibfnamefont{A.~E.} \bibnamefont{Nelson}},
  \bibinfo{journal}{JHEP} \textbf{\bibinfo{volume}{04}}, \bibinfo{pages}{170}
  (\bibinfo{year}{2015}), \eprint{1412.3113}.

\bibitem[{\citenamefont{Gonzalez~Macias and Wudka}(2015)}]{Macias:2015cna}
\bibinfo{author}{\bibfnamefont{V.}~\bibnamefont{Gonzalez~Macias}}
  \bibnamefont{and} \bibinfo{author}{\bibfnamefont{J.}~\bibnamefont{Wudka}},
  \bibinfo{journal}{JHEP} \textbf{\bibinfo{volume}{07}}, \bibinfo{pages}{161}
  (\bibinfo{year}{2015}), \eprint{1506.03825}.

\bibitem[{\citenamefont{González-Macías
  et~al.}(2016)\citenamefont{González-Macías, Illana, and
  Wudka}}]{Gonzalez-Macias:2016vxy}
\bibinfo{author}{\bibfnamefont{V.}~\bibnamefont{González-Macías}},
  \bibinfo{author}{\bibfnamefont{J.~I.} \bibnamefont{Illana}},
  \bibnamefont{and} \bibinfo{author}{\bibfnamefont{J.}~\bibnamefont{Wudka}},
  \bibinfo{journal}{JHEP} \textbf{\bibinfo{volume}{05}}, \bibinfo{pages}{171}
  (\bibinfo{year}{2016}), \eprint{1601.05051}.

\bibitem[{\citenamefont{Batell et~al.}(2017{\natexlab{a}})\citenamefont{Batell,
  Han, and Shams Es~Haghi}}]{Batell:2017rol}
\bibinfo{author}{\bibfnamefont{B.}~\bibnamefont{Batell}},
  \bibinfo{author}{\bibfnamefont{T.}~\bibnamefont{Han}}, \bibnamefont{and}
  \bibinfo{author}{\bibfnamefont{B.}~\bibnamefont{Shams Es~Haghi}}
  (\bibinfo{year}{2017}{\natexlab{a}}), \eprint{1704.08708}.

\bibitem[{\citenamefont{Batell et~al.}(2017{\natexlab{b}})\citenamefont{Batell,
  Han, McKeen, and Shams Es~Haghi}}]{Batell:2017cmf}
\bibinfo{author}{\bibfnamefont{B.}~\bibnamefont{Batell}},
  \bibinfo{author}{\bibfnamefont{T.}~\bibnamefont{Han}},
  \bibinfo{author}{\bibfnamefont{D.}~\bibnamefont{McKeen}}, \bibnamefont{and}
  \bibinfo{author}{\bibfnamefont{B.}~\bibnamefont{Shams Es~Haghi}}
  (\bibinfo{year}{2017}{\natexlab{b}}), \eprint{1709.07001}.

\bibitem[{\citenamefont{Pospelov}(2011)}]{Pospelov:2011ha}
\bibinfo{author}{\bibfnamefont{M.}~\bibnamefont{Pospelov}},
  \bibinfo{journal}{Phys. Rev.} \textbf{\bibinfo{volume}{D84}},
  \bibinfo{pages}{085008} (\bibinfo{year}{2011}), \eprint{1103.3261}.

\bibitem[{\citenamefont{Fox et~al.}(2011)\citenamefont{Fox, Liu, Tucker-Smith,
  and Weiner}}]{Fox:2011qd}
\bibinfo{author}{\bibfnamefont{P.~J.} \bibnamefont{Fox}},
  \bibinfo{author}{\bibfnamefont{J.}~\bibnamefont{Liu}},
  \bibinfo{author}{\bibfnamefont{D.}~\bibnamefont{Tucker-Smith}},
  \bibnamefont{and} \bibinfo{author}{\bibfnamefont{N.}~\bibnamefont{Weiner}},
  \bibinfo{journal}{Phys. Rev.} \textbf{\bibinfo{volume}{D84}},
  \bibinfo{pages}{115006} (\bibinfo{year}{2011}), \eprint{1104.4127}.

\bibitem[{\citenamefont{Chang and Weiner}(2008)}]{Chang:2007de}
\bibinfo{author}{\bibfnamefont{S.}~\bibnamefont{Chang}} \bibnamefont{and}
  \bibinfo{author}{\bibfnamefont{N.}~\bibnamefont{Weiner}},
  \bibinfo{journal}{JHEP} \textbf{\bibinfo{volume}{05}}, \bibinfo{pages}{074}
  (\bibinfo{year}{2008}), \eprint{0710.4591}.

\bibitem[{\citenamefont{Kopp and Welter}(2014)}]{Kopp:2014fha}
\bibinfo{author}{\bibfnamefont{J.}~\bibnamefont{Kopp}} \bibnamefont{and}
  \bibinfo{author}{\bibfnamefont{J.}~\bibnamefont{Welter}},
  \bibinfo{journal}{JHEP} \textbf{\bibinfo{volume}{12}}, \bibinfo{pages}{104}
  (\bibinfo{year}{2014}), \eprint{1408.0289}.

\bibitem[{\citenamefont{Chu et~al.}(2015)\citenamefont{Chu, Dasgupta, and
  Kopp}}]{Chu:2015ipa}
\bibinfo{author}{\bibfnamefont{X.}~\bibnamefont{Chu}},
  \bibinfo{author}{\bibfnamefont{B.}~\bibnamefont{Dasgupta}}, \bibnamefont{and}
  \bibinfo{author}{\bibfnamefont{J.}~\bibnamefont{Kopp}},
  \bibinfo{journal}{JCAP} \textbf{\bibinfo{volume}{1510}}, \bibinfo{pages}{011}
  (\bibinfo{year}{2015}), \eprint{1505.02795}.

\bibitem[{\citenamefont{Ng and Beacom}(2014)}]{Ng:2014pca}
\bibinfo{author}{\bibfnamefont{K.~C.~Y.} \bibnamefont{Ng}} \bibnamefont{and}
  \bibinfo{author}{\bibfnamefont{J.~F.} \bibnamefont{Beacom}},
  \bibinfo{journal}{Phys. Rev.} \textbf{\bibinfo{volume}{D90}},
  \bibinfo{pages}{065035} (\bibinfo{year}{2014}), \bibinfo{note}{[Erratum:
  Phys. Rev. D90, 089904 (2014)]}, \eprint{1404.2288}.

\bibitem[{\citenamefont{Pospelov and Pradler}(2012)}]{Pospelov:2012gm}
\bibinfo{author}{\bibfnamefont{M.}~\bibnamefont{Pospelov}} \bibnamefont{and}
  \bibinfo{author}{\bibfnamefont{J.}~\bibnamefont{Pradler}},
  \bibinfo{journal}{Phys. Rev.} \textbf{\bibinfo{volume}{D85}},
  \bibinfo{pages}{113016} (\bibinfo{year}{2012}), \bibinfo{note}{[Erratum:
  Phys. Rev.D88,no.3,039904(2013)]}, \eprint{1203.0545}.

\bibitem[{\citenamefont{Cherry et~al.}(2014)\citenamefont{Cherry, Friedland,
  and Shoemaker}}]{Cherry:2014xra}
\bibinfo{author}{\bibfnamefont{J.~F.} \bibnamefont{Cherry}},
  \bibinfo{author}{\bibfnamefont{A.}~\bibnamefont{Friedland}},
  \bibnamefont{and} \bibinfo{author}{\bibfnamefont{I.~M.}
  \bibnamefont{Shoemaker}} (\bibinfo{year}{2014}), \eprint{1411.1071}.

\bibitem[{\citenamefont{Berryman et~al.}(2017)\citenamefont{Berryman,
  de~Gouvêa, Kelly, and Zhang}}]{Berryman:2017twh}
\bibinfo{author}{\bibfnamefont{J.~M.} \bibnamefont{Berryman}},
  \bibinfo{author}{\bibfnamefont{A.}~\bibnamefont{de~Gouvêa}},
  \bibinfo{author}{\bibfnamefont{K.~J.} \bibnamefont{Kelly}}, \bibnamefont{and}
  \bibinfo{author}{\bibfnamefont{Y.}~\bibnamefont{Zhang}}
  (\bibinfo{year}{2017}), \eprint{1706.02722}.

\bibitem[{\citenamefont{Farzan and Heeck}(2016)}]{Farzan:2016wym}
\bibinfo{author}{\bibfnamefont{Y.}~\bibnamefont{Farzan}} \bibnamefont{and}
  \bibinfo{author}{\bibfnamefont{J.}~\bibnamefont{Heeck}},
  \bibinfo{journal}{Phys. Rev.} \textbf{\bibinfo{volume}{D94}},
  \bibinfo{pages}{053010} (\bibinfo{year}{2016}), \eprint{1607.07616}.

\bibitem[{\citenamefont{de~Gouvêa and Kobach}(2016)}]{deGouvea:2015euy}
\bibinfo{author}{\bibfnamefont{A.}~\bibnamefont{de~Gouvêa}} \bibnamefont{and}
  \bibinfo{author}{\bibfnamefont{A.}~\bibnamefont{Kobach}},
  \bibinfo{journal}{Phys. Rev.} \textbf{\bibinfo{volume}{D93}},
  \bibinfo{pages}{033005} (\bibinfo{year}{2016}), \eprint{1511.00683}.

\bibitem[{\citenamefont{Alekhin et~al.}(2016)}]{Alekhin:2015byh}
\bibinfo{author}{\bibfnamefont{S.}~\bibnamefont{Alekhin}} \bibnamefont{et~al.},
  \bibinfo{journal}{Rept. Prog. Phys.} \textbf{\bibinfo{volume}{79}},
  \bibinfo{pages}{124201} (\bibinfo{year}{2016}), \eprint{1504.04855}.

\bibitem[{\citenamefont{Bonivento et~al.}(2013)}]{Bonivento:2013jag}
\bibinfo{author}{\bibfnamefont{W.}~\bibnamefont{Bonivento}}
  \bibnamefont{et~al.} (\bibinfo{year}{2013}), \eprint{1310.1762}.

\bibitem[{\citenamefont{Adams et~al.}(2013)}]{Adams:2013qkq}
\bibinfo{author}{\bibfnamefont{C.}~\bibnamefont{Adams}} \bibnamefont{et~al.}
  (\bibinfo{collaboration}{LBNE}) (\bibinfo{year}{2013}), \eprint{1307.7335}.

\bibitem[{\citenamefont{Blondel et~al.}(2016)\citenamefont{Blondel, Graverini,
  Serra, and Shaposhnikov}}]{Blondel:2014bra}
\bibinfo{author}{\bibfnamefont{A.}~\bibnamefont{Blondel}},
  \bibinfo{author}{\bibfnamefont{E.}~\bibnamefont{Graverini}},
  \bibinfo{author}{\bibfnamefont{N.}~\bibnamefont{Serra}}, \bibnamefont{and}
  \bibinfo{author}{\bibfnamefont{M.}~\bibnamefont{Shaposhnikov}}
  (\bibinfo{collaboration}{FCC-ee study Team}), \bibinfo{journal}{Nucl. Part.
  Phys. Proc.} \textbf{\bibinfo{volume}{273-275}}, \bibinfo{pages}{1883}
  (\bibinfo{year}{2016}), \eprint{1411.5230}.

\bibitem[{\citenamefont{Deppisch et~al.}(2015)\citenamefont{Deppisch,
  Bhupal~Dev, and Pilaftsis}}]{Deppisch:2015qwa}
\bibinfo{author}{\bibfnamefont{F.~F.} \bibnamefont{Deppisch}},
  \bibinfo{author}{\bibfnamefont{P.~S.} \bibnamefont{Bhupal~Dev}},
  \bibnamefont{and}
  \bibinfo{author}{\bibfnamefont{A.}~\bibnamefont{Pilaftsis}},
  \bibinfo{journal}{New J. Phys.} \textbf{\bibinfo{volume}{17}},
  \bibinfo{pages}{075019} (\bibinfo{year}{2015}), \eprint{1502.06541}.

\bibitem[{\citenamefont{Dolgov}(2002)}]{Dolgov:2002wy}
\bibinfo{author}{\bibfnamefont{A.~D.} \bibnamefont{Dolgov}},
  \bibinfo{journal}{Phys. Rept.} \textbf{\bibinfo{volume}{370}},
  \bibinfo{pages}{333} (\bibinfo{year}{2002}), \eprint{hep-ph/0202122}.

\bibitem[{\citenamefont{Ade et~al.}(2016)}]{Ade:2015xua}
\bibinfo{author}{\bibfnamefont{P.~A.~R.} \bibnamefont{Ade}}
  \bibnamefont{et~al.} (\bibinfo{collaboration}{Planck}),
  \bibinfo{journal}{Astron. Astrophys.} \textbf{\bibinfo{volume}{594}},
  \bibinfo{pages}{A13} (\bibinfo{year}{2016}), \eprint{1502.01589}.

\bibitem[{\citenamefont{Bartoszek et~al.}(2014)}]{Bartoszek:2014mya}
\bibinfo{author}{\bibfnamefont{L.}~\bibnamefont{Bartoszek}}
  \bibnamefont{et~al.} (\bibinfo{collaboration}{Mu2e}) (\bibinfo{year}{2014}),
  \eprint{1501.05241}.

\bibitem[{\citenamefont{Adrian-Martinez
  et~al.}(2015)}]{Adrian-Martinez:2015wey}
\bibinfo{author}{\bibfnamefont{S.}~\bibnamefont{Adrian-Martinez}}
  \bibnamefont{et~al.} (\bibinfo{collaboration}{ANTARES}),
  \bibinfo{journal}{JCAP} \textbf{\bibinfo{volume}{1510}}, \bibinfo{pages}{068}
  (\bibinfo{year}{2015}), \eprint{1505.04866}.

\bibitem[{\citenamefont{Aartsen et~al.}(2016)}]{Aartsen:2016pfc}
\bibinfo{author}{\bibfnamefont{M.~G.} \bibnamefont{Aartsen}}
  \bibnamefont{et~al.} (\bibinfo{collaboration}{IceCube}),
  \bibinfo{journal}{Eur. Phys. J.} \textbf{\bibinfo{volume}{C76}},
  \bibinfo{pages}{531} (\bibinfo{year}{2016}), \eprint{1606.00209}.

\bibitem[{\citenamefont{Cohen et~al.}(2017)\citenamefont{Cohen, Murase, Rodd,
  Safdi, and Soreq}}]{Cohen:2016uyg}
\bibinfo{author}{\bibfnamefont{T.}~\bibnamefont{Cohen}},
  \bibinfo{author}{\bibfnamefont{K.}~\bibnamefont{Murase}},
  \bibinfo{author}{\bibfnamefont{N.~L.} \bibnamefont{Rodd}},
  \bibinfo{author}{\bibfnamefont{B.~R.} \bibnamefont{Safdi}}, \bibnamefont{and}
  \bibinfo{author}{\bibfnamefont{Y.}~\bibnamefont{Soreq}},
  \bibinfo{journal}{Phys. Rev. Lett.} \textbf{\bibinfo{volume}{119}},
  \bibinfo{pages}{021102} (\bibinfo{year}{2017}), \eprint{1612.05638}.

\bibitem[{\citenamefont{Weiler}(1999)}]{Weiler:1997sh}
\bibinfo{author}{\bibfnamefont{T.~J.} \bibnamefont{Weiler}},
  \bibinfo{journal}{Astropart. Phys.} \textbf{\bibinfo{volume}{11}},
  \bibinfo{pages}{303} (\bibinfo{year}{1999}), \eprint{hep-ph/9710431}.

\bibitem[{\citenamefont{Ioka and Murase}(2014)}]{Ioka:2014kca}
\bibinfo{author}{\bibfnamefont{K.}~\bibnamefont{Ioka}} \bibnamefont{and}
  \bibinfo{author}{\bibfnamefont{K.}~\bibnamefont{Murase}},
  \bibinfo{journal}{PTEP} \textbf{\bibinfo{volume}{2014}},
  \bibinfo{pages}{061E01} (\bibinfo{year}{2014}), \eprint{1404.2279}.

\bibitem[{\citenamefont{Spergel and Steinhardt}(2000)}]{Spergel:1999mh}
\bibinfo{author}{\bibfnamefont{D.~N.} \bibnamefont{Spergel}} \bibnamefont{and}
  \bibinfo{author}{\bibfnamefont{P.~J.} \bibnamefont{Steinhardt}},
  \bibinfo{journal}{Phys. Rev. Lett.} \textbf{\bibinfo{volume}{84}},
  \bibinfo{pages}{3760} (\bibinfo{year}{2000}), \eprint{astro-ph/9909386}.

\bibitem[{\citenamefont{Loeb and Weiner}(2011)}]{Loeb:2010gj}
\bibinfo{author}{\bibfnamefont{A.}~\bibnamefont{Loeb}} \bibnamefont{and}
  \bibinfo{author}{\bibfnamefont{N.}~\bibnamefont{Weiner}},
  \bibinfo{journal}{Phys. Rev. Lett.} \textbf{\bibinfo{volume}{106}},
  \bibinfo{pages}{171302} (\bibinfo{year}{2011}), \eprint{1011.6374}.

\bibitem[{\citenamefont{Tulin et~al.}(2013)\citenamefont{Tulin, Yu, and
  Zurek}}]{Tulin:2013teo}
\bibinfo{author}{\bibfnamefont{S.}~\bibnamefont{Tulin}},
  \bibinfo{author}{\bibfnamefont{H.-B.} \bibnamefont{Yu}}, \bibnamefont{and}
  \bibinfo{author}{\bibfnamefont{K.~M.} \bibnamefont{Zurek}},
  \bibinfo{journal}{Phys. Rev.} \textbf{\bibinfo{volume}{D87}},
  \bibinfo{pages}{115007} (\bibinfo{year}{2013}), \eprint{1302.3898}.

\bibitem[{\citenamefont{Kaplinghat et~al.}(2016)\citenamefont{Kaplinghat,
  Tulin, and Yu}}]{Kaplinghat:2015aga}
\bibinfo{author}{\bibfnamefont{M.}~\bibnamefont{Kaplinghat}},
  \bibinfo{author}{\bibfnamefont{S.}~\bibnamefont{Tulin}}, \bibnamefont{and}
  \bibinfo{author}{\bibfnamefont{H.-B.} \bibnamefont{Yu}},
  \bibinfo{journal}{Phys. Rev. Lett.} \textbf{\bibinfo{volume}{116}},
  \bibinfo{pages}{041302} (\bibinfo{year}{2016}), \eprint{1508.03339}.

\bibitem[{\citenamefont{Boehm et~al.}(2001)\citenamefont{Boehm, Fayet, and
  Schaeffer}}]{Boehm:2000gq}
\bibinfo{author}{\bibfnamefont{C.}~\bibnamefont{Boehm}},
  \bibinfo{author}{\bibfnamefont{P.}~\bibnamefont{Fayet}}, \bibnamefont{and}
  \bibinfo{author}{\bibfnamefont{R.}~\bibnamefont{Schaeffer}},
  \bibinfo{journal}{Phys. Lett.} \textbf{\bibinfo{volume}{B518}},
  \bibinfo{pages}{8} (\bibinfo{year}{2001}), \eprint{astro-ph/0012504}.

\bibitem[{\citenamefont{Buen-Abad et~al.}(2015)\citenamefont{Buen-Abad,
  Marques-Tavares, and Schmaltz}}]{Buen-Abad:2015ova}
\bibinfo{author}{\bibfnamefont{M.~A.} \bibnamefont{Buen-Abad}},
  \bibinfo{author}{\bibfnamefont{G.}~\bibnamefont{Marques-Tavares}},
  \bibnamefont{and} \bibinfo{author}{\bibfnamefont{M.}~\bibnamefont{Schmaltz}},
  \bibinfo{journal}{Phys. Rev.} \textbf{\bibinfo{volume}{D92}},
  \bibinfo{pages}{023531} (\bibinfo{year}{2015}), \eprint{1505.03542}.

\bibitem[{\citenamefont{Hooper and Zurek}(2008)}]{Hooper:2008im}
\bibinfo{author}{\bibfnamefont{D.}~\bibnamefont{Hooper}} \bibnamefont{and}
  \bibinfo{author}{\bibfnamefont{K.~M.} \bibnamefont{Zurek}},
  \bibinfo{journal}{Phys. Rev.} \textbf{\bibinfo{volume}{D77}},
  \bibinfo{pages}{087302} (\bibinfo{year}{2008}), \eprint{0801.3686}.

\bibitem[{\citenamefont{Arkani-Hamed and Weiner}(2008)}]{ArkaniHamed:2008qp}
\bibinfo{author}{\bibfnamefont{N.}~\bibnamefont{Arkani-Hamed}}
  \bibnamefont{and} \bibinfo{author}{\bibfnamefont{N.}~\bibnamefont{Weiner}},
  \bibinfo{journal}{JHEP} \textbf{\bibinfo{volume}{12}}, \bibinfo{pages}{104}
  (\bibinfo{year}{2008}), \eprint{0810.0714}.

\bibitem[{\citenamefont{Cheung et~al.}(2009)\citenamefont{Cheung, Ruderman,
  Wang, and Yavin}}]{Cheung:2009qd}
\bibinfo{author}{\bibfnamefont{C.}~\bibnamefont{Cheung}},
  \bibinfo{author}{\bibfnamefont{J.~T.} \bibnamefont{Ruderman}},
  \bibinfo{author}{\bibfnamefont{L.-T.} \bibnamefont{Wang}}, \bibnamefont{and}
  \bibinfo{author}{\bibfnamefont{I.}~\bibnamefont{Yavin}},
  \bibinfo{journal}{Phys. Rev.} \textbf{\bibinfo{volume}{D80}},
  \bibinfo{pages}{035008} (\bibinfo{year}{2009}), \eprint{0902.3246}.

\end{thebibliography}
\end{document}